\documentclass[12pt]{article}
\usepackage{amsmath}
\usepackage{graphicx,psfrag,epsf}
\usepackage{enumerate}
\usepackage{natbib}
\usepackage{url} % not crucial - just used below for the URL 
\usepackage[linesnumbered,ruled,lined]{algorithm2e}
\usepackage{algpseudocode}
\usepackage{multirow}
\usepackage{caption}
\usepackage{amsmath}

\usepackage{hyperref}
\hypersetup{
    colorlinks=true,
    linkcolor=blue,
    filecolor=blue,      
    urlcolor=cyan,
    citecolor=blue
    }
    
\usepackage{subcaption}

\newcommand{\blind}{0}

\begin{document}

\def\spacingset#1{\renewcommand{\baselinestretch}%
{#1}\small\normalsize} \spacingset{1}

%%%%%%%%%%%%%%%%%%%%%%%%%%%%%%%%%%%%%%%%%%%%%%%%%%%%%%%%%%%%%%%%%%%%%%%%%%%%%%

\if0\blind
{
  \title{\bf Generalized Variable Selection Algorithms for Gaussian Process Models by LASSO-like Penalty}
  \author{Zhiyong Hu\thanks{
    The authors report there are no competing interests to declare.}\hspace{.2cm}\\
    Department of Statistics, University of Connecticut\\
    and \\
    Dipak K. Dey \\
    Department of Statistics, University of Connecticut}
  \maketitle
} \fi

\if1\blind
{
  \title{\bf Generalized Variable Selection Algorithms for Gaussian Process Models by LASSO-like Penalty}
  
  \maketitle
} \fi

\bigskip
\begin{abstract}
With the rapid development of modern technology, massive amounts of data with complex pattern are generated. Gaussian process models that can easily fit the non-linearity in data become more and more popular nowadays. It is often the case that in some data only a few features are important or active. However, unlike classical linear models, it is challenging to identify active variables in Gaussian process models. One of the most commonly used methods for variable selection in Gaussian process models is automatic relevance determination, which is known to be open-ended. There is no rule of thumb to determine the threshold for dropping features, which makes the variable selection in Gaussian process models ambiguous. In this work, we propose two variable selection algorithms for Gaussian process models, which use the artificial nuisance columns as baseline for identifying the active features. Moreover, the proposed methods work for both regression and classification problems. The algorithms are demonstrated using comprehensive simulation experiments and an application to multi-subject electroencephalography data that studies alcoholic levels of experimental subjects.
\end{abstract}

\noindent%
\vfill

\newpage
\spacingset{1.45} % DON'T change the spacing!
\section{Introduction}
\label{sec:intro}
Most of the real world datasets are very complex, and it is hard to capture the complicated pattern of the data. Investigating such datasets requires flexible models that can learn the pattern or distribution of the data sufficiently. The parametric models, such as linear regression (LR) and generalized linear model (GLM), are often not able to detect the non-linearity or very hard to find the proper interactions in the data, which restricts the versatility of the model. As a result, the parametric model often becomes inadequate for solving complex real world problems. Thus, non-parametric methods like Gaussian process model that can be easily extended from LR or GLM become more attractive.

According to the definition in \cite{gpml}, a Gaussian process is a collection of random variables, any finite number of which have a joint Gaussian distribution, and it is completely specified by its mean function and covariance function: 
        \begin{equation}
f(\mathbf{x}) \sim \mathcal{G} \mathcal{P}\left(m(\mathbf{x}), C\left(\mathbf{x}, \mathbf{x}^{\prime}\right)\right)
\end{equation}
Conventionally, the mean function $m(\mathbf{x})$ is taken to be zero for simplicity. Given any finite $n$ and  $\mathbf{x}_1,\ldots,\mathbf{x}_n \in \mathcal{R}^K$, then $(f(\mathbf{x}_1),\ldots,f(\mathbf{x}_n))^\mathrm{T} \sim \mathcal{N}(\mathbf{0},\Sigma)$, where $\Sigma_{i,j} = C(\mathbf{x}_i,\mathbf{x}_j)$. Here, $C\left(\cdot, \cdot \right)$ is the covariance function. Radial basis function (RBF) is one of the most commonly used covariance functions in practice, which has form 
        \begin{equation} \label{eq:rbf}
C\left(\mathbf{x}_{i}, \mathbf{x}_{j}\right)=\sigma^{2} \exp \left(- \frac{1}{2} \sum_{k=1}^{K} \frac{\left(x_{i, k}-x_{j, k}\right)^{2}}{\gamma_{k}^{2}} \right),
\end{equation}
where marginal variance $\sigma^2$ and length-scales $\{ \gamma_{k}^{2} \}^K_{k=1}$ are the hyperparameters that control the shape of the Gaussian process. 

With the rapid development of modern technology, massive amounts of data are generated. However, it is common to encounter the case where data has high dimensional inputs but limited number of observations. Out of the large amount of inputs, it is often the case that only a few features are really important or active. Unlike the classical GLM that can select the variables using LASSO (e.g., \cite{tibshirani1996regression}, \cite{park2008bayesian}), it is challenging to identify active variables in Gaussian process models. 

\cite{gpml} pointed out that, for example, in covariance kernel (\ref{eq:rbf}) the inverse of length-scale $\gamma_{k}$ decides how relevant the $k_{th}$ feature is. As the length-scale $\gamma_{k}$ increases, the $k_{th}$ feature becomes less relevant. When $\gamma_{k}$ is estimated to be large, the $k_{th}$ feature is expected to have an ignorable impact on the covariance kernel, thus it can be removed from the model effectively. This is so-called automatic relevance determination (ARD) (e.g., \cite{williams1995gaussian}, \cite{neal2012bayesian}), which is one of the most commonly used methods for variable selection in Gaussian process models. However, it is known that ARD is open-ended, which does not have a clear rule for the threshold. More specifically, there is no rule of thumb to determine the threshold for dropping the features, which makes the variable selection in Gaussian process models ambiguous.

In this study, we introduce two algorithms that incorporate inverse-RBF kernel and LASSO-like regularizing exponential prior for detecting and selecting active variables in Gaussian process models, using different added nuisance columns as baseline. These algorithms are applicable in a general context. In addition to regression (i.e., Gaussian response), the proposed variable selection methods are also applicable for classification (e.g., Binomial or Poisson response) tasks, which have not been extensively studied in previous research. The performance of the algorithms is demonstrated using comprehensive simulation experiments and an application to multi-subject electroencephalography data that studies alcoholic levels of experimental subjects. The main focus of this manuscript is on variable selection in Gaussian process classification, and the appendix provides comparisons of our proposed method with the spike and slab variational Gaussian process (SSVGP) introduced by \cite{dance2022fast}, which demonstrate that our proposed method also performs adequately in regression tasks.

\subsection{Related Studies}
\medskip
Variable selection is a critical aspect of Gaussian process modeling. Many studies have been conducted on this topic, including \cite{oakley2004probabilistic}, \cite{schonlau2006screening}, and \cite{savitsky2011variable}. \cite{jiang2021variable} studied the consistency property of variable selection in squared exponential kernels. Recently, \cite{dance2022fast} introduced a variable selection method using the SSVGP, which demonstrated superior performance compared to benchmark algorithms such as LASSO. Despite the attractive performance of the SSVGP, it is not applicable for classification tasks.

\cite{linkletter2006variable} proposed a method for variable selection for Gaussian process models in computer experiments, which augments the design matrix $\mathrm{X}$ by including an inert variable and comparing it with the existing experimental variables. The Knockoffs approach for linear models proposed by \cite{barber2015controlling} also utilizes inert variables, but the Knockoffs method includes a dummy variable for each predictor in the original design matrix, resulting in a new design matrix with double the number of columns as the original. The method in \cite{linkletter2006variable} only considers the case where the response variable follows a normal distribution and Markov chain Monte Carlo (MCMC) is employed, which limits its generalizability. 

Unlike the prior choice in \cite{linkletter2006variable}, we utilize a LASSO-like regularizing exponential prior to place penalty on the inverse length-scales. \cite{gu2019jointly} showed that the LASSO penalty is a special class of jointly robust priors, which possess desirable characteristics. A R package that utilizes the joint robust prior for variable selection in Gaussian process regression was developed in \cite{gu2018robustgasp}, however variable selection for Gaussian process classification is still not addressed. 

The remaining part of this paper is organized as follows. Section \ref{sec:meth} provides details of the proposed methods. In Section \ref{sec:sim}, several simulation examples are demonstrated. Section \ref{sec:eeg} presents an application to multi-subject electroencephalography data. Finally, some discussions and concluding remarks are given in Section \ref{sec:discussion}.

\section{Methods}
\label{sec:meth}
\medskip
This section begins by reviewing the general Gaussian process models. For the dependent variables from a distribution in the exponential family, such as Gaussian, Poisson, and Bernoulli distributions, the general design behind Gaussian process models is straightforward, which is to assign a Gaussian process prior over the latent function, namely the systematic component in GLM. Therefore, the Gaussian process models have the following form:
\begin{equation}\label{eq:gpc}
\begin{array}{cl}
E(y_i|\mathbf{x}_i)= \mu_i = g^{-1}(f(\mathbf{x}_i)), \quad i=1,\ldots,n,\\
f(\cdot) \sim \mathcal{G} \mathcal{P}(\mathbf{0},C(\cdot,\cdot | \boldsymbol{\theta})),
\end{array}
\end{equation}\\
where $\mathbf{x}_i$ is the feature vector of the $i_{th}$ observation and $g$ is the link function. We specify different link functions according to the desired distribution for dependent variable $y$, for example, identity-link is used when the response is normally distributed, log-link is applied when the response follows Poisson distribution, and logit-link is the canonical link for Bernoulli variables. 

\subsection{Inverse-RBF Kernel}\label{sec: inverse rbf}
In practice, the RBF kernel (\ref{eq:rbf}) is one of the most widely-adopted kernels (see \cite{gpml}, \cite{stein2012interpolation}). However, if the RBF kernel is used for ARD, the interpretation of the length-scales is less straightforward. Moreover, it is hard to assign regularization to the length-scales since a larger length-scale indicates lower relevance.

Therefore, instead of the commonly used version of RBF kernel, we adopt the re-parameterized inverse-RBF kernel, which has this form:
\begin{equation} \label{eq:reverse_rbf}
C\left(\mathbf{x}_{i}, \mathbf{x}_{j}\right)=\sigma^{2} \exp \left(- \frac{1}{2} \sum_{k=1}^{K} \ell_{k}^{2} \times \left(x_{i, k}-x_{j, k}\right)^{2}\right),
\end{equation}
where the inverse length-scale $\ell
_k^2$ determines how relevant an input is. If the inverse length-scale $\ell_k^2$ has a large value, any difference in $k_{th}$ input would cause large impact on the covariance. Conversely, the input would only have ignorable influence on the covariance kernel if the inverse length-scale has a value that is close or even equal to 0, thus we can remove it from the model effectively. Therefore, the interpretation is similar to that of the coefficients of regular GLM models. 

In order to achieve the goal of selecting variables, regularization is needed to shrink the inverse length-scales of unnecessary features. In terms of Bayesian perspective, priors with large proportion of mass concentrated near 0 is needed for the inverse length-scales. \cite{park2008bayesian} discussed the Bayesian LASSO and demonstrated that the LASSO regression can be interpreted as a Bayesian regression with Laplace priors, which has density function proportional to
\begin{equation}
\pi(\beta) \propto  \exp \left(-\tau|\beta|\right),\quad \beta \in (-\infty, \infty).
\end{equation}
Since the inverse length-scale $\ell^2$ is non-negative, the Laplace prior is not applicable in this case, therefore we set the prior for $\ell^2$ as 
\begin{equation}
\pi(\ell^2) \propto
\left\{\begin{array}{l}
\exp(-\tau|\ell^2|) \quad \text{if} \quad \ell^2\geq 0 ;\\
0 \quad  \quad \quad  \quad \quad  \;\,\, \text{if} \quad  \ell^2< 0
\end{array} \right. \Rightarrow \pi(\ell^2) \propto \exp(-\tau\ell^2),
\end{equation}
which is actually an exponential distribution with mean $\frac{1}{\tau}$. The joint distribution of the model can be expressed as:
        \begin{equation}\label{eq:likelihood}
            \prod_{i = 1}^n p(y_i|f_i) 
            \times 
            \frac{1}{\sqrt{(2 \pi)^{n}|\mathbf{C}|}}\, \mathcal{}{e}^{ -\frac{1}{2} \mathbf{f}^{\mathrm{T}} \mathbf{C}^{-1}\mathbf{f}}
            \times
            \pi(\sigma^2)
            \times 
            \prod_{k=1}^K \pi(\ell_k^2),
        \end{equation}
where $\prod_{i = 1}^n p(y_i|f_i) $ is the model likelihood, $\mathbf{f}$ is the realization of Gaussian process, $\mathbf{C}$ is the corresponding covariance matrix, and $\pi(\sigma^2)$ represents the hyper-prior for marginal variance $\sigma^2$. Inserting the exponential priors for $\{\ell_{k}^{2} \}^K_{k=1}$ and taking logarithm of equation (\ref{eq:likelihood}), the log joint density of the model is proportional to
\begin{equation}\label{eq:log likelihood}
            \sum_{i = 1}^n \log p(y_i|f_i) 
            -\frac{1}{2}
            \log |\mathbf{C}|-\frac{1}{2}\mathbf{f}^{\mathrm{T}} \mathbf{C}^{-1}\mathbf{f}
            +
            \log \pi(\sigma^2)
            -
            \tau \sum_{k=1}^K  \ell_k^2.
        \end{equation}
From equation (\ref{eq:log likelihood}), it is obvious to see the last element $\tau \sum_{k=1}^K  \ell_k^2$ plays a role of regularization. That is, as the $\ell^2$ gets larger, it introduces more penalty to the log joint density. Also, by increasing the rate parameter $\tau$ for the exponential priors, the degree of penalty grows up. 

Since the posterior distributions are analytically intractable, the Bayesian computation is conducted for the model inference due to its flexibility. Rather than the computationally intensive MCMC methods, the variational inference (VI) approach that has been more and more popular recently for its computational advantages (see \cite{hensman2013gaussian}, \cite{hoffman2013stochastic}, and \cite{blei2017variational}) is employed in this work. 

Suppose $\theta$ is a latent variable of interest and $\mathbf{y}$ is the observed data, the basic idea about VI is to approximate the posterior distribution $p(\theta|\mathbf{y})$ by the chosen family of variational distribution $q(\theta)$, and then minimize the Kullback-Leibler (KL) divergence to the posterior distribution. Thus, it becomes an optimization problem and the goal is to search for the variational distribution $q^*(\theta)$ that minimize $KL(q(\theta)||p(\theta|\mathbf{y}))$. It turns out that minimizing $KL(q(\theta)||p(\theta|\mathbf{y}))$ is equivalent to maximize the evidence lower bound (ELBO) (see \cite{blei2017variational}), 
\begin{equation}\label{eq:elbo}
    \text{ELBO} = \text{E}_q\,[\log p(\theta, \mathbf{y})]-\text{E}_q\,[\log q(\theta)].
\end{equation}
Thus, the posterior distributions of the latent variables can be approximated using VI and the parameter estimates can be obtained, given the joint distribution (\ref{eq:likelihood}). \cite{hensman2013gaussian} and \cite{hensman2015scalable} discussed about the VI for Gaussian process models extensively. 

Nowadays, many packages for probabilistic modeling have implemented VI as a suggested inferential technique, such as but not limited to Edward (\cite{tran2016edward}) and Pyro (\cite{bingham2019pyro}). The algorithms in this study are developed on top of Pyro's framework to take advantage of its GPU acceleration and modeling flexibility. It is important to note that the algorithms proposed in this study do not impose specific requirements on the computational technique utilized, as long as it is able to effectively address the Bayesian Gaussian process problem. Thus, users are afforded the flexibility to utilize the technique with which they are most comfortable.

\subsection{Variable Selection Algorithms}
\medskip
As noted previously, ARD is open ended, which does not have a clear rule for the threshold. \cite{linkletter2006variable} proposed a method for variable selection for Gaussian process models in computer experiments, which augments the design matrix $\mathrm{X}$ by including an inert variable and compares it with the existing experimental variables. More specifically, consider the design matrix $\mathrm{X} = (x_1,x_2,\ldots,x_K)$, where $x_k$ is the column vector of $k_{th}$ feature, a nuisance vector $x_0$ that is irrelevant to the data is binded to it, then the augmented design matrix is $\mathrm{X^*} = (x_0, \mathrm{X})$. The feature $x_0$ is known to be irrelevant, therefore it is expected to have no impact on the prediction if it is included in the model. In other words, the corresponding inverse length-scale $\ell_0^2$ for $x_0$ should be very close to 0, then the researchers are confident to remove any inputs from the model if their inverse length-scales are even smaller than the $\ell_0^2$. However, this method is originally designed for normal response and only considers the case in which computationally expensive Markov chain Monte Carlo (MCMC) method is used. 

\begin{figure}
    \centering
    \includegraphics[scale = 0.6]{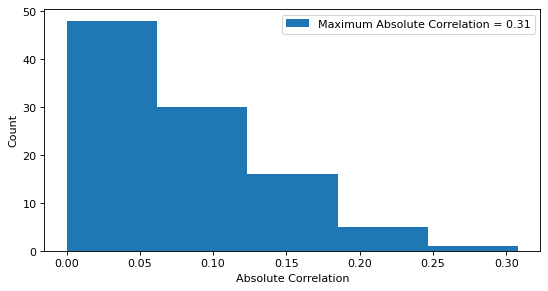}
    \caption{\small Randomly draw a sample of 100 standard normal variables and then draw other 100 samples from standard normal distribution, this figure shows the distribution of the absolute correlation between the first sample and last 100 samples. The maximum absolute correlation in this example is 0.31.}
    \label{fig:random_sample_hist}
\end{figure}

Ideally, the added nuisance column should be absolutely irrelevant to the data in order to eliminate any effect on the model. However, this is often not achievable when the sample size is small but the dimension is relatively high. For example, suppose we draw a sample of 100 standard normal variables as the nuisance variable $x_0$ and then draw other 100 samples from standard normal distribution as the original features $\mathrm{X}$ in data. The maximum absolute correlation between the first sample and last 100 samples can easily get close to 0.3, which is a relatively large value in practice. Figure \ref{fig:random_sample_hist} shows an example of histogram of absolute correlation between random samples as described previously. 

To address this issue, the data augmentation should be repeated several times, say $\mathcal{M}$ times, and record the parameter estimates for the inverse length-scales of $x_0$ and all columns in $\mathrm{X}$ at each time. By repeating the procedure, it averages over the effect of the inert columns. As for the parameter estimates, we utilize the maximum a posteriori probability (MAP) estimates obtained from the VI discussed in Section \ref{sec: inverse rbf}, since the model is under a Bayesian setup. After obtaining the parameter estimates, we compare the distribution of these estimates to select active variables.

Denote $\hat{\ell}^2_{j} = (\hat{\ell}^2_{1, j},\ldots, \hat{\ell}^2_{m, j},\ldots,\hat{\ell}^2_{\mathcal{M}, j})^{\mathrm{T}}$, where $\hat{\ell}^2_{m, j}$ is the MAP estimate for the inverse length-scale of $x_j$ from the VI in the $m_{th}$ iteration of the algorithm, $j = 0, 1,\ldots, K$. For each $j \in \{1,\ldots,K\}$, we compare the distribution of $\hat{\ell}^2_{j}$ with the distribution of $\hat{\ell}^2_{0}$ to evaluate whether the $j_{th}$ feature is active or not. More specifically, the $q_{th}$ percentile of $\hat{\ell}^2_{0}$, denoted as $\alpha_q$, is used as the threshold for selecting active inputs. That is, if the median of $\hat{\ell}^2_{j}$ is smaller than $\alpha_q$, then the $j_{th}$ feature $x_j$ is considered inactive, thus it can be dropped from the model effectively. 

To make the inverse length-scales comparable, the design matrix $\mathrm{X}$ needs to be standardized at first such that each column has zero mean and unit standard deviation. Algorithm \ref{alg:random} summarizes the steps for variable selection with random nuisance columns. 

\begin{algorithm}[t]
\caption{Variable Selection with Random Nuisance Columns}\label{alg:random}
\KwData{$\mathrm{X}, \mathbf{y}$}
\KwIn{$\mathcal{M}$, $q$}
standardize $\mathrm{X}$\;
\For {$m \in \{1,\ldots,\mathcal{M}\}$}{
augment $\mathrm{X}$ by a random vector $x_0$ from the design space, s.t. $\mathrm{X^*} = (x_0, \mathrm{X})$\;
obtain and record MAP estimates for all inverse length-scales using VI\;
}

store $\mathcal{L} = (\hat{\ell}^2_{0},\hat{\ell}^2_{1},\ldots,\hat{\ell}^2_{K})$ and calculate $\alpha_q $: the $ q_{th}$ percentile of $\hat{\ell}^2_{0}$\;
\For {$k \in \{1,\ldots,K\}$}{
  \eIf{median($\hat{\ell}^2_{k}$)$\geq \alpha_q$}{
    $x_k$ is active;
  }{
    $x_k$ is inactive.
  }
}
\KwOut{Index of active features}

\end{algorithm}

The way $q$ is specified indicates how much the researcher is willing to accept the identification of an inactive feature as active. Hence, the researcher also has the option to set distinct thresholds for individual features. Choosing a larger value of $q$ indicates that the researcher prioritizes a lower rate of mistakenly identifying inactive inputs as active. However, it will increase the risk of failing to detect some weakly active inputs. From the empirical results, an intuitive explanation of the threshold is to consider $(100-q)\%$ as the maximum acceptable rate of falsely identifying inactive inputs as active. The simulation experiments, which are detailed in this paper and the corresponding appendix, consistently demonstrate rates of mistakenly identifying inactive inputs as active that are either close to or lower than $(100-q)\%$.

In practice, $q$ should be at least 50 since the active feature should have a larger impact on the model than the known nuisance feature, which only have minimal influence on the model (i.e., $\hat{\ell}^2_{0}$ should stay close to 0). As for how to set an optimal threshold, we would recommend to conduct grid search for the threshold and penalty level of the prior, like how users search for the regularizing strength parameter $\lambda$ when using LASSO. 

As for the number of iterations $\mathcal{M}$, a larger value is always preferred since more iterations of augmentation can average down the effect of the nuisance columns. However, $\mathcal{M}$ is restricted by the computational power and computing time. Though the recent development of Bayesian computation like variational inference has accelerated the model-fitting process significantly, the Gaussian process model is still very computationally intensive for its $\mathcal{O}(n^3)$ complexity. By our empirical research, it turns out that in general $\mathcal{M}=20$ iterations are enough for identifying active features accurately. 

\begin{algorithm}[t]
\caption{Variable Selection with PCA Transformed Columns}\label{alg:pca}
\KwData{$\mathrm{X}, \mathbf{y}$}
\KwIn{$\mathcal{M}$, $q$}
standardize $\mathrm{X}$\;
run PCA transformation on $\mathrm{X}$ to get $\mathrm{X}_{PCA}$\;
standardize $\mathrm{X}_{PCA}$\;
\For {$m \in \{1,\ldots,\mathcal{M}\}$}{
augment $\mathrm{X}$ by $m_{th}$ from the last column of $\mathrm{X}_{PCA}$\;
obtain and record MAP estimates for all inverse length-scales using VI\;
}

store $\mathcal{L} = (\hat{\ell}^2_{0},\hat{\ell}^2_{1},\ldots,\hat{\ell}^2_{K})$ and calculate $\alpha_q $: the $ q_{th}$ percentile of $\hat{\ell}^2_{0}$\;
\For {$k \in \{1,\ldots,K\}$}{
  \eIf{median($\hat{\ell}^2_{k}$)$\geq \alpha_q$}{
    $x_k$ is active;
  }{
    $x_k$ is inactive.
  }
}
\KwOut{Index of active features}
\end{algorithm}

Besides randomly sample variables from the design space as the nuisance column, we propose to conduct principal component analysis (PCA) on design matrix $\mathrm{X}$ and use the last few PCA transformed columns as the nuisance columns for variable selection. It is known that the first few principle components explain most of the variance in the data, while the last few principle components are often dropped. Thus, the last few PCA transformed columns can be considered nuisance and used in the variable selection algorithm. Another advantage of using PCA transformed columns is that PCA is an orthogonal linear transformation such that transformed data are uncorrelated, thus we can ensure that the added column in each iteration is uncorrelated to others. Algorithm \ref{alg:pca} summarizes the steps for variable selection with PCA transformed columns. 

\begin{figure}[t]
    \centering
    \includegraphics[scale = 0.6]{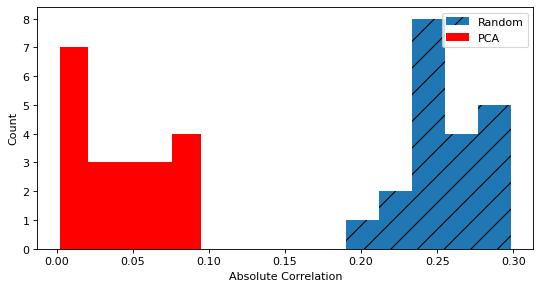}
    \caption{\small The part with hatch shows the histogram of maximum absolute correlation when 20 random nuisance columns are used, while the part without hatch shows the histogram of maximum absolute correlation when the last 20 PCA transformed columns of $\mathrm{X}$ are used.}
    \label{fig:pca_random}
\end{figure}

Next, a simple example is illustrated. Suppose we draw a $100 \times 100$ matrix of standard normal variables as the original design matrix $\mathrm{X}$ (i.e., $n=100, K=100$), then draw a sample of 100 standard normal variables as $x_0$ and record the maximum absolute correlation between $x_0$ and $\mathrm{X}$, 20 times. This shows a simple example of what the correlation between added nuisance column and the original features is like in Algorithm \ref{alg:random}. The part with hatch of the Figure \ref{fig:pca_random} shows the histogram of maximum absolute correlation of the aforementioned case. Instead, the part without hatch of Figure \ref{fig:pca_random} shows the histogram of maximum absolute correlation when the last 20 PCA transformed columns of $\mathrm{X}$ are used as nuisance columns accordingly. It is clear that the maximum absolute correlation is lower on average when the PCA transformed columns are used as nuisance columns in this example.

However, it is obvious that when the dimension of data is relatively low even the last few PCA transformed columns are likely to be correlated to the original $\mathrm{X}$, also the range of $\mathcal{M}$ is limited. Thus, the Algorithm \ref{alg:pca} is not recommended when the number of features in data is relatively small.

\section{Simulations}\label{sec:sim}
\medskip
To demonstrate the performance of the proposed algorithms in different scenarios, several simulation experiments are conducted. All the dependent variables of simulated data are generated from Bernoulli distribution in this manuscript, and simulation experiments with Gaussian dependent variables are presented in the appendix. For the marginal variance parameter $\sigma^2$, a log-normal $\mathcal{LN}(0,2)$ prior is specified in all applications. 
\subsection{Example 1}
\medskip
The design matrix $\mathrm{X}$ of the first example is generated by drawing a $100 \times 71$ matrix (i.e., $n=100,K=71$) of standard normal variables, but only the first feature of the data is actually active. In this example, we are simulating the scenario where we only have a small number of observations but a relatively large number of features. The latent probability of the dependent variable is determined by
\begin{equation}
    E(y_i|\mathbf{x}_i)= \mu_i = g^{-1}(3\sin{x_{i, 1}}), \quad i=1,\ldots,100,
\end{equation}
where $g(\cdot)$ is the logit-link, $\mathbf{x}_i$ is the feature vector of the $i_{th}$ observation, and $x_{i, k}$ represents the $k_{th}$ element in $\mathbf{x}_{i}$. With the simulated data, we apply both algorithms to identify the active feature. For all the applications in this work, the number of iterations $\mathcal{M}$ is set to be 20. Exp(4) prior for the inverse length-scales is used in this example.

    \begin{figure*}
        \centering
        \begin{subfigure}[b]{0.48\textwidth}
            \centering
            \includegraphics[width=\textwidth]{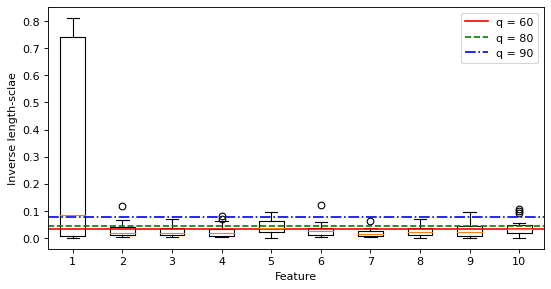}
            \caption[]%
            {{\small Random nuisance columns}}    
            \label{fig: ex1random}
        \end{subfigure}
        \hfill
        \begin{subfigure}[b]{0.48\textwidth}  
            \centering 
            \includegraphics[width=\textwidth]{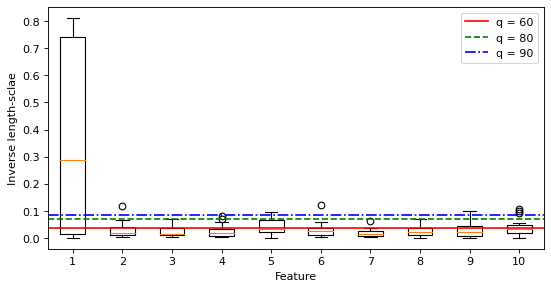}
            \caption[]%
            {{\small PCA transformed columns}}    
            \label{fig: ex1pca}
        \end{subfigure}
        \hfill

        \caption{\small Box-plots of the first 10 features of $\mathcal{L}$ obtained from (a) Algorithm \ref{alg:random} and (b) Algorithm \ref{alg:pca} in example 1. The horizontal solid, dashed, and dash-dot lines represent $\alpha_{60}$, $\alpha_{80}$ and $\alpha_{90}$ respectively.}
        \label{fig: ex1}
    \end{figure*}

For visual illustration, Figure \ref{fig: ex1} shows the Box-plots of the first 10 features of $\mathcal{L}$, while other 61 features are ignored due to the limited space. The horizontal solid, dashed, and dash-dot represent $\alpha_{60}$, $\alpha_{80}$ and $\alpha_{90}$ respectively. The graphical form of the algorithm outcome makes it convenient for researchers to check if a feature is active. Notice that all the thresholding lines stay close to 0 as expected, indicating that the added nuisance variable has small influence on the model.

By simply comparing the median lines of the corresponding Box-plots and the dashed lines in Figure \ref{fig: ex1random}, we can determine the first factor is active, though it is very marginal when $q = 90$. As noted, the specification of $q$ reflects the researcher's tolerance of false identification of an inactive feature as active. As the $q$ increases, the number of false discovery decreases. As for Algorithm \ref{alg:pca}, from Figure \ref{fig: ex1pca}, the median of the $\hat{\ell}^2_{1}$ is even much larger than $\alpha_{90}$, which gives us more confidence that the first feature is active. 
% \usepackage{multirow}
\begin{table}
\centering
\caption{\small Table shows the rate of being identified as active by different algorithms in 50 repetitions of simulations of example 1.}\label{tab:sim1}
\begin{tabular}{|c|c|cc|}
\hline
\multirow{2}{*}{Threshold} & \multirow{2}{*}{Algorithm} & \multicolumn{2}{c|}{Proportion of Discovery} \\ \cline{3-4} 
                           &                            & Inactive Features       & $x_1$      \\ \hline
\multirow{2}{*}{$\alpha_{60}$}        & Random                     & 0.04                    & 1         \\
                           & PCA                        & 0.04                    & 0.98      \\ \hline
\multirow{2}{*}{$\alpha_{80}$}        & Random                     & \textless{}0.01         & 0.94      \\
                           & PCA                        & \textless{}0.01         & 0.96      \\ \hline
\multirow{2}{*}{$\alpha_{90}$}        & Random                     & \textless{}0.01         & 0.88      \\
                           & PCA                        & \textless{}0.01         & 0.92      \\ \hline
\end{tabular}
\end{table}

The simulation is repeated 50 times to check the performance of both algorithms. Table \ref{tab:sim1} shows the proportion of discovery, the proportion of times a feature is identified as active, of the actual active feature $x_1$ and all other 70 inactive features using thresholds $\alpha_{60}$, $\alpha_{80}$, and $\alpha_{90}$. Both algorithms work well as they can identify the active feature correctly in most of the cases. The rate of falsely identifying an inactive feature as active is very low. However, as the $q$ increases, the cut-off becomes more conservative, thus the proportion of discovery decreases. The researchers should pay attention to the choice of $q$ in practice. Overall, the Algorithm \ref{alg:pca} performs slightly better than the other since it correctly detects the true active feature more often when the threshold is conservative. 

\subsection{Example 2}
\medskip
The design matrix $\mathrm{X}$ of the second example is generated by drawing a $500 \times 56$ matrix (i.e., $n=500,K=56$) of standard normal variables, and the latent probability of the dependent variable is determined by
\begin{equation}\label{eq:sim2}
    E(y_i|\mathbf{x}_i)= \mu_i = g^{-1}(x_{i, 1}-x_{i, 2}-\frac{1}{2} x_{i, 3}+\frac{1}{4} x_{i, 4}+ \frac{1}{8}x_{i, 5}+\frac{1}{16} x_{i, 6}),\quad i=1,\ldots,500,
\end{equation}
where $g(\cdot)$ is the logit-link, $\mathbf{x}_i$ is the feature vector of the $i_{th}$ observation, and $x_{i, k}$ represents the $k_{th}$ element in $\mathbf{x}_{i}$. Therefore, in this simulation we have 6 active or weakly active features and 50 absolutely inactive features. The format of the linear function is designed to test the sensitivity of the algorithms. The elements with smaller magnitude of coefficient tend to be less active.

% \usepackage{multirow}
\begin{table}
\centering
\caption{\small Table shows the rate of being identified as active by Algorithms \ref{alg:random} with different priors on $\ell^2$ in 50 repetitions of simulations of example 2.}\label{tab:sim2 prior}
\begin{tabular}{|c|c|ccccccc|}
\hline
\multirow{2}{*}{Threshold} & \multirow{2}{*}{Prior} & \multicolumn{7}{c|}{Proportion of Discovery}                                 \\ \cline{3-9} 
                           &                        & Inactive Features         & $x_1$   & $x_2$   & $x_3$   & $x_4$  & $x_5$   & $x_6$   \\ \hline
\multirow{5}{*}{$\alpha_{80}$}       & No Prior               & \multicolumn{1}{c|}{0.17} & 0.14 & 0.26 & 0.12 & 0.12 & 0.18  & 0.18 \\
                           & Exp(1)                 & \multicolumn{1}{c|}{0.10} & 1    & 1    & 0.94 & 0.62 & 0.20  & 0.04 \\
                           & Exp(2)                 & \multicolumn{1}{c|}{0.06} & 1    & 1    & 0.96 & 0.72 & 0.22 & 0.04 \\
                           & Exp(4)              & \multicolumn{1}{c|}{0.05} & 1    & 1    & 0.98 & 0.68 & 0.24 & 0.04 \\
                           & Exp(10)                & \multicolumn{1}{c|}{0.03} & 1    & 1    & 0.96 & 0.58 & 0.18 & 0.02 \\ \hline
\end{tabular}
\end{table}

The power of the regularization of the exponential prior used in the proposed algorithms is discussed at first. The simulation is repeated 50 times with different exponential priors on the inverse length-scales. We also run the algorithms without regularization (i.e., no prior on inverse length-scales) to show the importance of the proposed exponential priors. 

Table \ref{tab:sim2 prior} shows the proportion of discovery of the actual active or weakly active features and all other 50 inactive features of Algorithms \ref{alg:random} with different priors on $\ell^2$ in 50 repetitions of simulations. The threshold is fixed at $\alpha_{80}$. From the table, it is clear that the variable selection algorithm is not able to detect the active features when no regularization is assigned on the inverse length-scales. This is because the Gaussian process models can easily over-fit the data since the dimension is relatively high, thus even the estimate for inverse length-scale of the known nuisance vector can stay far away from 0. Instead, with the proposed exponential prior, the algorithm performs adequately. As noted, by increasing the rate parameter $\tau$ for the exponential prior, the degree of penalty increases. From Table \ref{tab:sim2 prior}, it is easy to notice that the proportion of times that inactive features are detected to be active decreases as the rate parameter for the exponential prior increases. However, the weakly active features are less likely to be detected if the penalty is too large. For example, the $4_{th}$ feature is less often identified as active when Exp(10) prior is used compared to others. In general, moderate regularization (e.g., Exp(2) or Exp(4) in this case) is recommended.

% \usepackage{multirow}
\begin{table}
\centering
\caption{\small Table shows the rate of being identified as active by different algorithms in 50 repetitions of simulations of example 2 using the Exp(4) prior.}\label{tab:sim2}
\begin{tabular}{|c|c|ccccccc|}
\hline
\multirow{2}{*}{Threshold} & \multirow{2}{*}{Algorithm} & \multicolumn{7}{c|}{Proportion of Discovery}                             \\ \cline{3-9} 
                           &                            & Inactive Features         & $x_1$ & $x_2$ & $x_3$   & $x_4$   & $x_5$   & $x_6$   \\ \hline
\multirow{2}{*}{$\alpha_{60}$}       & Random                     & \multicolumn{1}{c|}{0.35} & 1  & 1  & 1    & 0.86 & 0.58 & 0.22 \\
                           & PCA                        & \multicolumn{1}{c|}{0.38} & 1  & 1  & 1    & 0.94 & 0.62 & 0.16 \\ \hline
\multirow{2}{*}{$\alpha_{80}$}       & Random                     & \multicolumn{1}{c|}{0.05} & 1  & 1  & 0.98 & 0.68 & 0.24 & 0.04 \\
                           & PCA                        & \multicolumn{1}{c|}{0.08} & 1  & 1  & 0.96 & 0.72 & 0.28 & 0.06 \\ \hline
\multirow{2}{*}{$\alpha_{90}$}       & Random                     & \multicolumn{1}{c|}{0.01} & 1  & 1  & 0.92 & 0.48 & 0.12 & 0.02 \\
                           & PCA                        & \multicolumn{1}{c|}{0.01} & 1  & 1  & 0.90 & 0.50 & 0.14 & 0.02 \\ \hline
\end{tabular}
\end{table}

Table \ref{tab:sim2} shows the rate of identifying the features as active using the Exp(4) prior for inverse length-scales, under different thresholds from the two proposed algorithms. Both algorithms can easily identify the first 3 strongly active features. Algorithm \ref{alg:pca} performs slightly better on identifying the $4_{th}$ feature. Moreover, the $5_{th}$ feature, which has small coefficient, although it is hard to be detected, it is still identified more often than the absolutely inactive features. As for the $6_{th}$ feature, since its coefficient is too small the algorithm considers it as nuisance at most of the times.

\begin{figure*}
        \centering
        \begin{subfigure}[b]{0.48\textwidth}
            \centering
            \includegraphics[width=\textwidth]{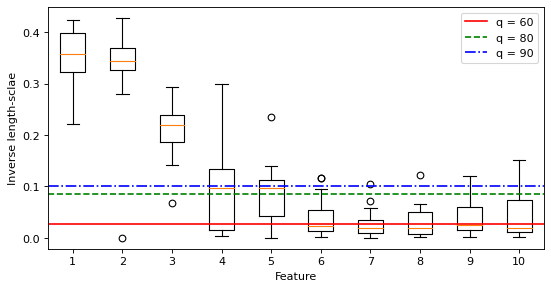}
            \caption[]%
            {{\small Random nuisance columns}}    
            \label{fig: sim2random}
        \end{subfigure}
        \hfill
        \begin{subfigure}[b]{0.48\textwidth}  
            \centering 
            \includegraphics[width=\textwidth]{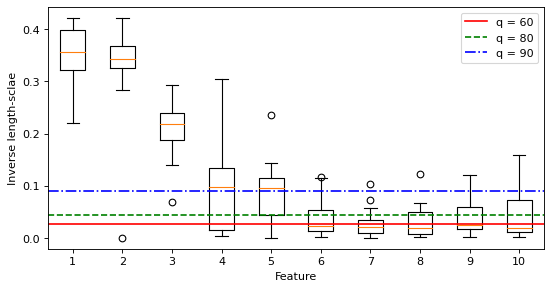}
            \caption[]%
            {{\small PCA transformed columns}}    
            \label{fig: sim2pca}
        \end{subfigure}
        \hfill

        \caption{\small Box-plots of the first 10 features of $\mathcal{L}$ obtained from (a) Algorithm \ref{alg:random} and (b) Algorithm \ref{alg:pca} in example 2. The horizontal solid, dashed, and dash-dot lines represent $\alpha_{60}$, $\alpha_{80}$ and $\alpha_{90}$ respectively.}
        \label{fig: ex2}
    \end{figure*}

Figure \ref{fig: ex2} shows the Box-plots of the first 10 features of $\mathcal{L}$ in one repetition of simulations in example 2, while other 46 features are ignored due to the limited space. The first 2 features are most active, thus their boxes locate above all others obviously. As the magnitude of the $3_{rd}$ to $5_{th}$ features' coefficients decreases, the corresponding boxes locate lower and lower. As for the $6_{th}$ feature, it is similar to those of inactive features as its coefficient is too small.

Next, we consider the case where the design matrix $\mathrm{X}$ contains binary columns. The latent probability of the dependent variables is still determined by equation (\ref{eq:sim2}). In this case, for each row $i$, $x_{i, 1}$ and $x_{i, 2}$ are drawn from Bernoulli distribution with success probability $0.5$, while the rest of the active predictors are still from standard normal distribution. Out of the 50 inactive predictors, 10 columns are drawn from Bernoulli distribution with success probability $0.5$, while the rest are drawn from standard normal distribution. 

Table \ref{tab:sim2 binary} shows the proportion of discovery of the features using the Exp(2) prior for inverse length-scales when binary columns exist, under different thresholds of Algorithm \ref{alg:pca}. The results indicate that the proposed algorithm performs adequately when the design matrix includes binary columns. Simulation experiments that sample all the columns of the design matrix from the Bernoulli distribution in the appendix also demonstrate the performance of the proposed method. 

\begin{table}
\centering
\caption{\small Table shows the rate of being identified as active by Algorithm \ref{alg:pca} with Exp(2) prior on $\ell^2$ in 50 repetitions of simulations of modified example 2 with binary predictors.}\label{tab:sim2 binary}
\begin{tabular}{|c|ccccccc|}
\hline
 \multirow{2}{*}{Threshold} & \multicolumn{7}{c|}{Proportion of Discovery}                                 \\ \cline{2-8} 
                           &               Inactive Features         & $x_1$   & $x_2$   & $x_3$   & $x_4$  & $x_5$   & $x_6$   \\ \hline
 $\alpha_{60}$               & \multicolumn{1}{c|}{0.27} & 0.98 & 1 & 1 & 0.88 & 0.66  & 0.26 \\
                           $\alpha_{80}$                 & \multicolumn{1}{c|}{0.03} & 0.94    & 1    & 0.98 & 0.72 & 0.18  & 0.06 \\
                          $\alpha_{90}$                 & \multicolumn{1}{c|}{0.01} & 0.90    & 0.98    & 0.94 & 0.42 & 0.08 & 0.02 \\ \hline
\end{tabular}
\end{table}

So far, the simulation experiments have only examined the case in which all predictors are independently drawn. In the next step, we will modify the design of this example to assess the performance of the proposed method on correlated data. Again, the latent probability of the dependent variables is determined by equation (\ref{eq:sim2}). But in this case, the predictors are correlated. Following the design of correlated features in \cite{barber2015controlling}, each row of the design matrix is drawn from a $\mathcal{N}(\mathbf{0},\Theta)$ distribution, where $\Theta_{i,i} = 1$ for all $i$ and $\Theta_{i,j} = 0.3$ for all $i \neq j$. 

Table \ref{tab:sim2 correlate} shows the proportion of discovery of the active or weakly active features $x_1,\ldots,x_6$ and all other inactive features using the Exp(2) prior for inverse length-scales when all features are correlated, under different thresholds of Algorithm \ref{alg:pca}. From the results, it is easy to observe that the performance of the proposed algorithm is only degraded on the weakly active features. Despite slightly inferior performance compared to the case in which all predictors are independent, the results suggest that the proposed algorithm performs satisfactorily even when all the features are correlated.

\begin{table}
\centering
\caption{\small Table shows the rate of being identified as active by Algorithm \ref{alg:pca} with Exp(2) prior on $\ell^2$ in 50 repetitions of simulations of modified example 2 with correlated predictors.}\label{tab:sim2 correlate}
\begin{tabular}{|c|ccccccc|}
\hline
\multirow{2}{*}{Threshold} & \multicolumn{7}{c|}{Proportion of Discovery}                                 \\ \cline{2-8} 
                           &            Inactive Features         & $x_1$   & $x_2$   & $x_3$   & $x_4$  & $x_5$   & $x_6$   \\ \hline
 $\alpha_{60}$               & \multicolumn{1}{c|}{0.37} & 1 & 1 & 1 & 0.94 & 0.58  & 0.26 \\
 $\alpha_{80}$                 & \multicolumn{1}{c|}{0.07} & 1    & 1    & 1 & 0.64 & 0.12  & 0.08 \\
 $\alpha_{90}$                 & \multicolumn{1}{c|}{0.02} & 1    & 1    & 0.98 & 0.42 & 0.06 & 0.02 \\ \hline
\end{tabular}
\end{table}

Moreover, we also checked if adding a nuisance column would have big influence on the estimates for other features. We repeat the algorithm without the nuisance columns in an iteration of simulation and record the $\mathcal{L}$ without $\hat{\ell}^2_{0}$. The average squared differences of estimated inverse length-scales between it and the two proposed algorithms are both only about $2.6\times10^{-3}$. Also, from the Figure \ref{fig: ex2} it is easy to see that the distributions of MAP estimates are very similar even the added nuisance columns are different in the two algorithms, suggesting that adding a nuisance column would not have large impacts on the estimates for other features.

\subsection{Example 3}
\medskip
In this example, a more complex non-linear latent probability function is considered. The design matrix $\mathrm{X}$ of this example is generated by drawing a $500 \times 54$ matrix (i.e., $n=500,K=54$) of standard normal variables, and the latent probability of the dependent variable is determined by
\begin{equation}\label{eq:sim3}
    E(y_i|\mathbf{x}_i)= \mu_i = g^{-1}(\sin{x_{i, 1}} + \frac{3}{2} \cos{x_{i, 2}} + 2 \sin{x_{i, 3}}+\frac{5}{2} \cos{x_{i, 4}}),\quad i=1,\ldots,500,
\end{equation}
where $g(\cdot)$ is the logit-link, $\mathbf{x}_i$ is the feature vector of the $i_{th}$ observation, and $x_{i, k}$ represents the $k_{th}$ element in $\mathbf{x}_{i}$. In this simulation, we have 4 active features and 50 absolutely inactive features. 

Similar to the previous examples, the simulation is repeated 50 times. In this example, Exp(2) and Exp(4) priors for inverse length-scales are used. Table \ref{tab:sim3} shows the proportion of discovery of the active features $x_1,\ldots,x_4$ and all other 50 inactive features under different thresholds. The active features are effectively detected by both algorithms. From Table \ref{tab:sim3}, it is also clear that the Exp(2) prior works better than the other one, suggesting that a better choice of prior on inverse length-scales can lead to better variable selection. In general, this example demonstrates that both algorithms work well on identifying active features, even the data has complicated non-linear structure. 

% \usepackage{multirow}
\begin{table}
\centering
\caption{\small Table shows the rate of being identified as active by different algorithms in 50 repetitions of simulations of example 3 with different priors on inverse length-scales.}\label{tab:sim3}
\begin{tabular}{|c|c|c|ccccc|}
\hline
\multirow{2}{*}{Prior}  & \multirow{2}{*}{Threshold} & \multirow{2}{*}{Algorithm} & \multicolumn{5}{c|}{Proportion of Discovery}               \\ \cline{4-8} 
                        &                            &                            & Inactive Features         & $x_1$   & $x_2$   & $x_3$ & $x_4$ \\ \hline
\multirow{6}{*}{Exp(2)} & \multirow{2}{*}{$\alpha_{60}$}       & Random                     & \multicolumn{1}{c|}{0.33} & 0.88 & 0.98 & 1  & 1  \\
                        &                            & PCA                        & \multicolumn{1}{c|}{0.37} & 0.90 & 1    & 1  & 1  \\ \cline{2-8} 
                        & \multirow{2}{*}{$\alpha_{80}$}       & Random                     & \multicolumn{1}{c|}{0.06} & 0.82 & 0.94 & 1  & 1  \\
                        &                            & PCA                        & \multicolumn{1}{c|}{0.07} & 0.80 & 0.94 & 1  & 1  \\ \cline{2-8} 
                        & \multirow{2}{*}{$\alpha_{90}$}       & Random                     & \multicolumn{1}{c|}{0.01} & 0.76 & 0.90 & 1  & 1  \\
                        &                            & PCA                        & \multicolumn{1}{c|}{0.01} & 0.70 & 0.94 & 1  & 1  \\ \hline
\multirow{6}{*}{Exp(4)} & \multirow{2}{*}{$\alpha_{60}$}       & Random                     & \multicolumn{1}{c|}{0.32} & 0.86 & 0.92 & 1  & 1  \\
                        &                            & PCA                        & \multicolumn{1}{c|}{0.41} & 0.86 & 0.92 & 1  & 1  \\ \cline{2-8} 
                        & \multirow{2}{*}{$\alpha_{80}$}      & Random                     & \multicolumn{1}{c|}{0.05} & 0.80 & 0.76 & 1  & 1  \\
                        &                            & PCA                        & \multicolumn{1}{c|}{0.07} & 0.78 & 0.78 & 1  & 1  \\ \cline{2-8} 
                        & \multirow{2}{*}{$\alpha_{90}$}       & Random                     & \multicolumn{1}{c|}{0.01} & 0.74 & 0.68 & 1  & 1  \\
                        &                            & PCA                        & \multicolumn{1}{c|}{0.01} & 0.74 & 0.68 & 1  & 1  \\ \hline
\end{tabular}
\end{table}

Similar to example 2, we adjust the design of this example to evaluate the performance of the proposed method on correlated data. The latent probability of the dependent variables is still determined by equation (\ref{eq:sim3}), but the predictors are correlated in this design. Again, for each row of the design matrix, $\mathbf{x}_{i}$ is drawn from a $\mathcal{N}(\mathbf{0},\Theta)$ distribution, where $\Theta_{i,i} = 1$ for all $i$ and $\Theta_{i,j} = 0.3$ for all $i \neq j$. Table \ref{tab:sim3 correlate} shows the proportion of discovery of the active features $x_1,\ldots,x_4$ and all other 50 inactive features using the Exp(2) prior for inverse length-scales, under different thresholds of Algorithm \ref{alg:pca}. Again, the proposed algorithm exhibits satisfactory performance even when all the features are correlated, although its performance is slightly inferior to the case in which all predictors are independent, as indicated by the results of simulation experiments.

\begin{table}
\centering
\caption{\small Table shows the rate of being identified as active by Algorithm \ref{alg:pca} with Exp(2) prior on $\ell^2$ in 50 repetitions of simulations of modified example 3 with correlated predictors.}\label{tab:sim3 correlate}
\begin{tabular}{|c|ccccc|}
\hline
\multirow{2}{*}{Threshold} & \multicolumn{5}{c|}{Proportion of Discovery}                                 \\ \cline{2-6} 
                                                   & Inactive Features         & $x_1$   & $x_2$   & $x_3$   & $x_4$    \\ \hline
 $\alpha_{60}$               & \multicolumn{1}{c|}{0.36} & 0.90 & 0.96 & 1 & 1  \\
                            $\alpha_{80}$                 & \multicolumn{1}{c|}{0.07} & 0.74    & 0.90    & 1 & 1  \\
                            $\alpha_{90}$                 & \multicolumn{1}{c|}{0.02} & 0.64    & 0.82    & 1 & 1  \\ \hline
\end{tabular}
\end{table}

\section{Application to Electroencephalography Data}\label{sec:eeg}
\medskip
In this section, we investigate a multi-subject electroencephalography (EEG) data, which measures voltage fluctuations resulting from ionic current within the neurons of the brain. Clinically, EEG refers to the recording of the brain's spontaneous electrical activity over a period of time, as recorded from multiple electrodes placed on the scalp (see \cite{eeg}). Thus, the EEG data is usually a 3-dimensional tensor that has dimension of $n \times K \times t$, where $n$, $K$ and $t$ are the number of experimental subjects, the number of electrodes (locations) and the number of time points respectively. The data used in our work is from an experiment on studying the EEG correlates of genetic predisposition to alcoholism (see \cite{hu2015local} and \cite{mohammed2019bayesian}) with dimension $122\times 57 \times 256$ (i.e., $n=122,K=57,t=256$). The experimental subjects are divided into two groups with 77 subjects in the alcoholic group and the other 45 subjects in the control group. During the experiment, stimulus was applied to each subject and the electrical activity is recorded. The target is to predict the alcoholic status of the subject given the EEG records.

Since the number of features $57\times 256$ is extremely high, it is not suitable to use general GLM in this case. Thus, we adopt the local aggregate modeling approach that fits local models at each time point separately proposed by \cite{mohammed2019bayesian}. The brain activities are known to be very complex, while the linear latent functions can hardly represent this level of complexity, therefore Gaussian process models are believed to be more appropriate in this case. At each time point, given the $122 \times 57$ design matrix, a Gaussian process model is fitted as the local model. However, at a particular time point it is clear that not all the regions of brain are activated. Learning which regions of brain are correlated with the stimulus is also a crucial topic. Thus, in addition to the local Gaussian process model (LGP), variable selection is also conducted to find the active locations at each time point using the proposed algorithms. For notational simplicity, the LGP incorporated with Algorithm \ref{alg:random} is denoted as LGP.1, while the LGP incorporated with Algorithm \ref{alg:pca} is denoted as LGP.2. In this work, both LGP.1 and LGP.2 use Exp(2) prior for the inverse length-scales.

As for the prediction for subject's alcoholic status, we predict the alcoholic status using each local model and record the responses sequentially through all time points. Thus, for each individual, we will have a binary prediction vector with length $256$. As suggested in \cite{mohammed2019bayesian}, we predict the subject level responses as the class indicator with the longest length of run.

We use 5-fold cross-validation to assess the performance of the out-of-fold prediction accuracy of different methods and the procedure is repeated several times for robust results. \cite{mohammed2019bayesian} proposed a local Bayesian model (LBM) with independent spike-and-slab prior for this problem. \cite{mohammed2020classification} then update the LBM using structured spike-and-slab prior that utilizes spatial information. The results from these two methods are also included for comparison.

\begin{table}
\centering
\caption{\small Table shows the average prediction accuracy and standard error across multiple 5-fold cross-validations for the EEG data from different methods.}\label{tab:eeg}
\begin{tabular}{|c|cc|}
\hline
Method             & Accuracy & Std. Err. \\ \hline
LBM (Independent)   & 0.701    & 0.040     \\
LBM (Structured)   & 0.717    & 0.029     \\
LGP.1 with $\alpha_{50}$ & 0.734    & 0.021     \\
LGP.1 with $\alpha_{55}$ & 0.727    & 0.024     \\
LGP.2 with $\alpha_{50}$ & 0.746    & 0.018     \\
LGP.2 with $\alpha_{55}$ & 0.729    & 0.012    \\ \hline
\end{tabular}
\end{table}

Table \ref{tab:eeg} shows the average prediction accuracy and standard error across multiple 5-fold cross-validations for the EEG data using different methods, from which we can easily see that the LGP methods are more accurate than the LBM methods. Moreover, the standard errors of LGP methods are also lower than LBM, which suggests our proposed models are more stable. As for the comparison among LGP methods, the LGP.2 shows higher accuracy in prediction with lower standard errors. Unlike the simulation experiments that have known highly active features, a threshold larger than $\alpha_{60}$ would over-sparsify the models for the EEG data, resulting poor estimations. The distribution of the number of EEG activated locations under different thresholds in one cross-validation is provided in the appendix. 

Though the prediction accuracy using threshold $\alpha_{55}$ is slightly lower compared to $\alpha_{50}$, it provides more reasonably sparse models. Therefore, we discuss the LGP models with $\alpha_{55}$ in the remaining part. 

    \begin{figure*}
        \centering
        \begin{subfigure}[b]{0.8\textwidth}
            \centering
            \includegraphics[width=\textwidth]{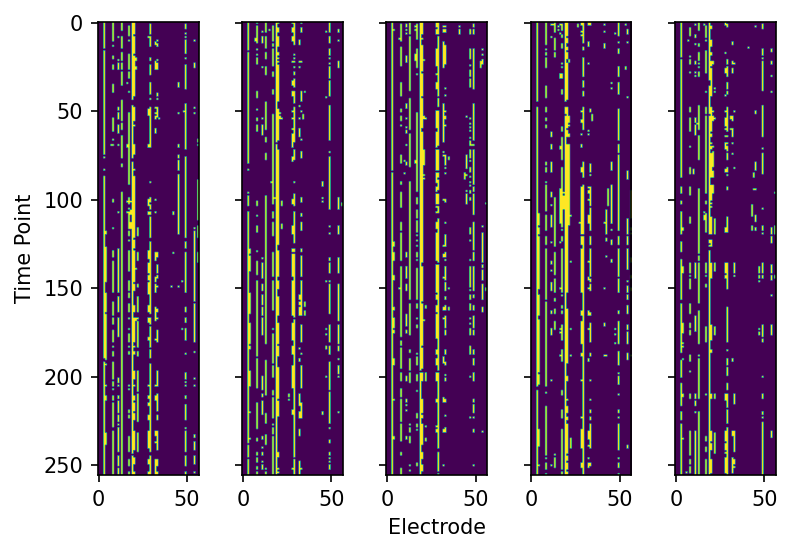}
            \caption[]%
            {{\small LGP.1 with $\alpha_{55}$}}    
            \label{fig: eeg_random}
        \end{subfigure}
        \vfill
        \begin{subfigure}[b]{0.8\textwidth}  
            \centering 
            \includegraphics[width=\textwidth]{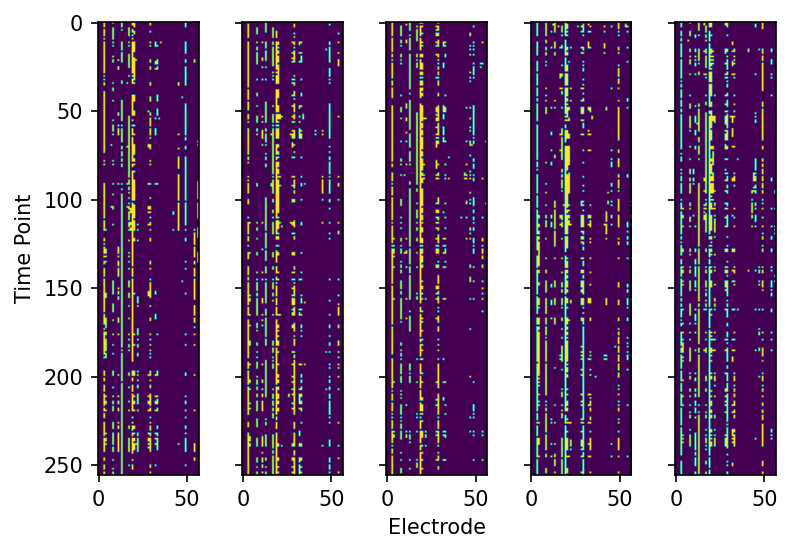}
            \caption[]%
            {{\small LGP.2 with $\alpha_{55}$}}    
            \label{fig: eeg_pca}
        \end{subfigure}
        \hfill

        \caption{\small EEG activated locations obtained from (a) LGP.1 with $\alpha_{55}$ and (b) LGP.2 with $\alpha_{55}$ in a 5-fold cross-validation. The black area indicates that the location of electrode at corresponding time point is not considered active by the algorithm. The plot from left to right in a sub-figure is obtained from cross-validation fold 1 to 5 respectively.}
        \label{fig: eeg}
    \end{figure*}

As noted, besides the prediction accuracy, identifying locations of brain that are correlated with the stimulus is also of interest. Figure \ref{fig: eeg} shows EEG activated locations across time that obtained from LGP methods with $\alpha_{55}$ in a 5-fold cross-validation. The black area indicates that the location of electrode at corresponding time point is not considered active by the algorithm. The plot from left to right in a sub-figure is obtained from cross-validation fold 1 to 5 respectively. From the figure, we notice that activated locations tend to stay active for a long time, which suggests the particular regions of brain are reacting to the stimulus consistently. Select a specific sub-plot in a sub-figure, by comparing the colored area with other sub-plots horizontally, we can check whether the variable selection algorithm is robust to identify the true active locations for different cohort of subjects, since the activated area for different individuals should be similar. According to Figure \ref{fig: eeg}, the identified active regions of both methods are consistent across different folds of subjects. By comparing the colored area with other sub-plot vertically, it is easy to check whether the two variable selection algorithms are identifying the similar group of locations. Although the detected active location sets of Algorithm \ref{alg:pca} are more sparse, the major part of the selected locations is quite similar in Figure \ref{fig: eeg_random} and Figure \ref{fig: eeg_pca}. This is a strong signal that both algorithms can identify the active regions consistently. The EEG activated locations obtained using $\alpha_{50}$ in the same 5-fold cross-validation is provided in the appendix for reference.

In summary, both LGP.1 and LGP.2 work well for the EEG data in predicting alcoholic status and identifying the active regions reacted to the stimulus. However, the LGP.2 has higher prediction accuracy with more sparse local models, which suggests that the Algorithm \ref{alg:pca} detects the active locations more rigorously for the EEG data.

\section{Discussion}\label{sec:discussion}
\medskip
This paper introduces two generalized variable selection algorithms for both Gaussian process regression and classification tasks, which use artificial nuisance columns as baseline for identifying the active features. We also propose to use inverse-RBF kernel and LASSO-like regularizing exponential prior on inverse length-scale parameters. The first algorithm uses completely random vector from the design space as the added inert column, while the second algorithm utilizes the least important PCA transformed columns as the nuisance columns since the last few PCA transformed columns only carry very few information. Moreover, another advantage of using the PCA transformed columns is that PCA is an orthogonal linear transformation such that transformed data are uncorrelated. However, the PCA transformed columns are not recommended when the dimension is relatively low. The simulation experiments and application to EEG data demonstrate the performance of the proposed algorithms. In particular, Algorithm \ref{alg:pca} shows promising capability of identifying sparse active features while keeping the important information from the EEG data.

Although both variable selection algorithms work well, they still suffer from the extremely time-consuming model-fitting operation of Gaussian process models. The time complexity of a Gaussian process is $\mathcal{O}(n^3)$, where $n$ is the sample size. Therefore, both proposed algorithms have $\mathcal{O}(\mathcal{M}n^3)$ complexity for identifying the active features. The complexity may be prohibitive when the sample size is large. Here, in the big data scenario, we suggest to adopt the variational sparse Gaussian process models proposed in \cite{hensman2013gaussian} and \cite{hensman2015scalable} which only have $\mathcal{O}(nS^2)$ complexity by utilizing $S$ inducing inputs, where $S$ is much smaller than $n$ (e.g., $S=n/10$). As a result, the complexity of the variable selection algorithms can be reduced to $\mathcal{O}(n\mathcal{M}S^2)$. Furthermore, when the computational resource is sufficient, the algorithms can be parallelized such that the model-fitting steps are conducted simultaneously on different augmented data. Nonetheless, it is still worth developing some alternative approaches that can reduce the value of $\mathcal{M}$ while keeping accurate detection with limited computational power. 

\bibliographystyle{JASA}
\bibliography{manuscript}

\begin{thebibliography}{25}
\newcommand{\enquote}[1]{``#1''}
\expandafter\ifx\csname natexlab\endcsname\relax\def\natexlab#1{#1}\fi

\bibitem[\protect\citename{Barber and Cand{\`e}s, }2015]{barber2015controlling}
Barber, Rina~Foygel and Cand{\`e}s, Emmanuel~J (2015), ``Controlling the false
  discovery rate via knockoffs,'' {\em The Annals of Statistics\/}, 43, 5,
  2055--2085.

\bibitem[\protect\citename{Bingham et~al., }2019]{bingham2019pyro}
Bingham, Eli, Chen, Jonathan~P, Jankowiak, Martin, Obermeyer, Fritz, Pradhan,
  Neeraj, Karaletsos, Theofanis, Singh, Rohit, Szerlip, Paul, Horsfall, Paul,
  and Goodman, Noah~D (2019), ``Pyro: Deep universal probabilistic
  programming,'' {\em The Journal of Machine Learning Research\/}, 20, 1,
  973--978.

\bibitem[\protect\citename{Blei et~al., }2017]{blei2017variational}
Blei, David~M, Kucukelbir, Alp, and McAuliffe, Jon~D (2017), ``Variational
  inference: A review for statisticians,'' {\em Journal of the American
  statistical Association\/}, 112, 518, 859--877.

\bibitem[\protect\citename{Dance and Paige, }2022]{dance2022fast}
Dance, Hugh and Paige, Brooks (2022),, Fast and Scalable Spike and Slab
  Variable Selection in High-Dimensional Gaussian Processes.
\newblock in {\em International Conference on Artificial Intelligence and
  Statistics\/},  7976--8002. PMLR.

\bibitem[\protect\citename{Gu, }2019]{gu2019jointly}
Gu, Mengyang (2019), ``Jointly robust prior for Gaussian stochastic process in
  emulation, calibration and variable selection,'' {\em Bayesian Analysis\/},
  14, 3, 857--885.

\bibitem[\protect\citename{Gu et~al., }2019]{gu2018robustgasp}
Gu, Mengyang, Palomo, Jesus, and Berger, James~O. (2019), ``{RobustGaSP: Robust
  Gaussian Stochastic Process Emulation in R},'' {\em {The R Journal}\/}, 11,
  1, 112--136.

\bibitem[\protect\citename{Hensman et~al., }2013]{hensman2013gaussian}
Hensman, James, Fusi, Nicolo, and Lawrence, Neil~D (2013), ``Gaussian processes
  for big data,'' {\em arXiv preprint arXiv:1309.6835\/}.

\bibitem[\protect\citename{Hensman et~al., }2015]{hensman2015scalable}
Hensman, James, Matthews, Alexander, and Ghahramani, Zoubin (2015),, Scalable
  variational Gaussian process classification.
\newblock in {\em Artificial Intelligence and Statistics\/},  351--360. PMLR.

\bibitem[\protect\citename{Hoffman et~al., }2013]{hoffman2013stochastic}
Hoffman, Matthew~D, Blei, David~M, Wang, Chong, and Paisley, John (2013),
  ``Stochastic variational inference.,'' {\em Journal of Machine Learning
  Research\/}, 14, 5.

\bibitem[\protect\citename{Hu and Allen, }2015]{hu2015local}
Hu, Yue and Allen, Genevera~I (2015), ``Local-aggregate modeling for big data
  via distributed optimization: Applications to neuroimaging,'' {\em
  Biometrics\/}, 71, 4, 905--917.

\bibitem[\protect\citename{Jiang and Tokdar, }2021]{jiang2021variable}
Jiang, Sheng and Tokdar, Surya~T (2021), ``Variable selection consistency of
  Gaussian process regression,'' {\em The Annals of Statistics\/}, 49, 5,
  2491--2505.

\bibitem[\protect\citename{Linkletter et~al., }2006]{linkletter2006variable}
Linkletter, Crystal, Bingham, Derek, Hengartner, Nicholas, Higdon, David, and
  Ye, Kenny~Q (2006), ``Variable selection for Gaussian process models in
  computer experiments,'' {\em Technometrics\/}, 48, 4, 478--490.

\bibitem[\protect\citename{Mohammed and Dey, }2020]{mohammed2020classification}
Mohammed, Shariq and Dey, Dipak~K (2020), ``Classification of high-dimensional
  electroencephalography data with location selection using structured
  spike-and-slab prior,'' {\em Statistical Analysis and Data Mining: The ASA
  Data Science Journal\/}, 13, 5, 465--481.

\bibitem[\protect\citename{Mohammed et~al., }2019]{mohammed2019bayesian}
Mohammed, Shariq, Dey, Dipak~K, and Zhang, Yuping (2019), ``Bayesian variable
  selection using spike-and-slab priors with application to high dimensional
  electroencephalography data by local modelling,'' {\em Journal of the Royal
  Statistical Society: Series C (Applied Statistics)\/}, 68, 5, 1305--1326.

\bibitem[\protect\citename{Neal, }2012]{neal2012bayesian}
Neal, Radford~M (2012), {\em Bayesian learning for neural networks\/}, vol.
  118, Springer Science \& Business Media.

\bibitem[\protect\citename{Niedermeyer and da~Silva, }2005]{eeg}
Niedermeyer, Ernst and da~Silva, FH~Lopes (2005), {\em Electroencephalography:
  basic principles, clinical applications, and related fields\/}, Lippincott
  Williams \& Wilkins.

\bibitem[\protect\citename{Oakley and O'Hagan, }2004]{oakley2004probabilistic}
Oakley, Jeremy~E and O'Hagan, Anthony (2004), ``Probabilistic sensitivity
  analysis of complex models: a Bayesian approach,'' {\em Journal of the Royal
  Statistical Society: Series B (Statistical Methodology)\/}, 66, 3, 751--769.

\bibitem[\protect\citename{Park and Casella, }2008]{park2008bayesian}
Park, Trevor and Casella, George (2008), ``The Bayesian LASSO,'' {\em Journal
  of the American Statistical Association\/}, 103, 482, 681--686.

\bibitem[\protect\citename{Rasmussen and Williams, }2006]{gpml}
Rasmussen, Carl~Edward and Williams, Christopher K.~I. (2006), {\em Gaussian
  processes for machine learning.\/}, Adaptive computation and machine
  learning, MIT Press.

\bibitem[\protect\citename{Savitsky et~al., }2011]{savitsky2011variable}
Savitsky, Terrance, Vannucci, Marina, and Sha, Naijun (2011), ``Variable
  selection for nonparametric Gaussian process priors: Models and computational
  strategies,'' {\em Statistical Science\/},  130--149.

\bibitem[\protect\citename{Schonlau and Welch, }2006]{schonlau2006screening}
Schonlau, Matthias and Welch, William~J (2006), ``Screening the input variables
  to a computer model via analysis of variance and visualization,'' In {\em
  Screening\/}, ed. I.~{\em Screening\/}, Springer,  pp.  308--327.

\bibitem[\protect\citename{Stein, }2012]{stein2012interpolation}
Stein, Michael~L (2012), {\em Interpolation of spatial data: some theory for
  kriging\/}, Springer Science \& Business Media.

\bibitem[\protect\citename{Tibshirani, }1996]{tibshirani1996regression}
Tibshirani, Robert (1996), ``Regression shrinkage and selection via the
  lasso,'' {\em Journal of the Royal Statistical Society: Series B
  (Methodological)\/}, 58, 1, 267--288.

\bibitem[\protect\citename{Tran et~al., }2016]{tran2016edward}
Tran, Dustin, Kucukelbir, Alp, Dieng, Adji~B., Rudolph, Maja, Liang, Dawen, and
  Blei, David~M. (2016), ``{Edward: A library for probabilistic modeling,
  inference, and criticism},'' {\em arXiv preprint arXiv:1610.09787\/}.

\bibitem[\protect\citename{Williams and Rasmussen, }1995]{williams1995gaussian}
Williams, Christopher~KI and Rasmussen, Carl~Edward (1995),, Gaussian processes
  for regression.
\newblock in {\em Proceedings of the 8th International Conference on Neural
  Information Processing Systems\/},  514--520.

\end{thebibliography}

\end{document}

\def\spacingset#1{\renewcommand{\baselinestretch}%
{#1}\small\normalsize} \spacingset{1}

%%%%%%%%%%%%%%%%%%%%%%%%%%%%%%%%%%%%%%%%%%%%%%%%%%%%%%%%%%%%%%%%%%%%%%%%%%%%%%

\if0\blind
{
  \title{\bf Generalized Variable Selection Algorithms for Gaussian Process Models by LASSO-like Penalty}
  \author{Zhiyong Hu\thanks{
    The authors report there are no competing interests to declare.}\hspace{.2cm}\\
    Department of Statistics, University of Connecticut\\
    and \\
    Dipak K. Dey \\
    Department of Statistics, University of Connecticut}
  \maketitle
} \fi

\if1\blind
{
  \title{\bf Generalized Variable Selection Algorithms for Gaussian Process Models by LASSO-like Penalty}
  
  \maketitle
} \fi

\spacingset{1.45} % DON'T change the spacing!

\section{Appendix}
\medskip
\subsection{Simulation Experiments with Binary Response Variables}
\medskip

\subsubsection{Simulation 1}\label{sec: inactive simulation}
\medskip
To check the performance of the proposed variable selection algorithm in the case that no predictor in the data is active, a simulation is conducted. In this simulation, the design matrix $\mathrm{X}$ is generated by drawing a $500 \times 50$ matrix (i.e., $n=500,K=50$) of standard normal variables, while the dependent variable is randomly drawn from Bernoulli distribution with success probability 0.5. Therefore, the dependent variable is irrelevant to the design matrix $\mathrm{X}$. 

Similar to the examples in the manuscript, the simulation is repeated 50 times.  Table \ref{tab:appendix sim 1} shows the proportion of discovery of the inactive features from Algorithm 2 with $Exp(2)$ prior for inverse length-scales. From the table, the proportion of false identification of the simulation experiments stay below $(100-q)\%$, indicating that the algorithm can control the proportion of false identification even when no predictors are active. 
\begin{table}
\centering
\caption{\small Table shows the rate of being identified as active by Algorithm 2 with $Exp(2)$ prior on $\ell^2$ in 50 repetitions of simulations.}\label{tab:appendix sim 1}
% \usepackage{multirow}
% \usepackage{multirow}
\begin{tabular}{|c|c|c|}
\hline
\multirow{2}{*}{Prior} & \multirow{2}{*}{Threshold} & Proportion of Discovery    \\ \cline{3-3} 
                       &                            & Inactive Features \\ \hline
\multirow{4}{*}{Exp(2)} & $\alpha_{60}$                         & 0.35              \\ \cline{2-3} 
                       & $\alpha_{70}$                         & 0.21              \\ \cline{2-3} 
                       & $\alpha_{80}$                         & 0.07              \\ \cline{2-3} 
                       & $\alpha_{90}$                         & 0.01              \\ \hline
\end{tabular}
\end{table}

\subsection{Simulation Experiments with Gaussian Response Variables}\label{sec: gaussian simulation}
\medskip
To demonstrate the generalizability of the proposed method, simulation experiments with Gaussian response variables are also conducted. In this Gaussian setup, we compare the variable selection performance of our proposed algorithm with the spike and slab variational Gaussian process (SSVGP) proposed by \cite{dance2022fast}. The implementation of SSVGP is from the Github provided by the authors using posterior inclusion probability (PIP) threshold 0.5. 

\subsubsection{Simulation 2}
\medskip
The design matrix $\mathrm{X}$ is generated by drawing a $500 \times 56$ matrix (i.e., $n=500,K=56$) of standard normal variables, the mean of the dependent variable is determined by 
\begin{equation}\label{eq:sim2 mean}
    E(y_i|\mathbf{x}_i)= \mu_i = x_{i, 1}-x_{i, 2}-\frac{1}{2} x_{i, 3}+\frac{1}{4} x_{i, 4}+ \frac{1}{8}x_{i, 5}+\frac{1}{16} x_{i, 6},\quad i=1,\ldots,500,
\end{equation}
where $\mathbf{x}_i$ is the feature vector of the $i_{th}$ observation, and $x_{i, k}$ represents the $k_{th}$ element in $\mathbf{x}_{i}$. Then the dependent variable $y_i$ is sampled from $\mathcal{N}(\mu_i, 1)$. Like the example 2 in the paper, we have 6 active or weakly active features and 50 absolutely inactive features in this simulation. The format of the linear function is designed to test the sensitivity of the algorithms. The elements with smaller magnitude of coefficient tend to be less active. 

Like the previous example, the simulation is repeated 50 times. Table \ref{tab:appendix sim 2} shows the proportion of discovery of the active or weakly active features $x_1,\ldots,x_6$, and the inactive features from SSVGP and Algorithm 2 with Exp(2) prior for inverse length-scales, from which we can notice that our proposed method performs comparably with the SSVGP in this simulation. Compared to the proposed method, the SSVGP tends to over-select weakly active features, leading to higher rate of false identification. 

\begin{table}
\centering
\caption{\small Table shows the rate of being identified as active by different algorithms in 50 repetitions of simulations with Gaussian responses.}\label{tab:appendix sim 2}
% \usepackage{multirow}
\begin{tabular}{|c|c|ccccccc|}
\hline
\multirow{2}{*}{Prior} & \multirow{2}{*}{Algorithm} & \multicolumn{7}{c|}{Proportion of Discovery}                             \\ \cline{3-9} 
                       &                            & Inactive Features         & $x_1$   & $x_2$   & $x_3$   & $x_4$  & $x_5$ & $x_6$   \\ \hline
                       & SSVGP                      & \multicolumn{1}{c|}{0.17} & 1  & 1  & 1    & 1    & 0.70 & 0.36 \\ \hline
\multirow{2}{*}{Exp(2)} & Algorithm 2 with $\alpha_{80}$ & \multicolumn{1}{c|}{0.15} & 1  & 1  & 1    & 0.94 & 0.62 & 0.24 \\ \cline{2-9} 
                       & Algorithm 2 with $\alpha_{90}$ & \multicolumn{1}{c|}{0.05} & 1  & 1  & 0.98 & 0.78 & 0.26 & 0.08 \\ \hline
\end{tabular}
\end{table}

Next, we consider the case in which all the columns of the design matrix are drawn from Bernoulli distribution, meaning that all the features are binary. For each column in $\mathrm{X}$, the success probability $p$ is first sampled from uniform distribution $\mathcal{U}(0.1, 0.9)$, then the entries of this column are sampled from Bernoulli distribution with success probability $p$. The mean of the dependent variables is still determined by equation (\ref{eq:sim2 mean}). Table \ref{tab:appendix sim 2 binary} shows the proportion of discovery from SSVGP and Algorithm 2 with Exp(2) prior of 50 repetitions of simulation. As indicated by the results, the proposed algorithm with $\alpha_{80}$ shows comparable performance to the SSVGP, despite the tendency of the SSVGP to over-select weakly active features. Also, the $\alpha_{90}$ is too conservative in this case. It is observed that the performance of variable selection for both algorithms is slightly diminished when utilizing binary features, as indicated by a comparison of Table \ref{tab:appendix sim 2} and Table \ref{tab:appendix sim 2 binary}. This degradation in performance is likely due to the presence of sharp transitions between the two states of binary features, which negatively impacts the performance of the inverse-RBF kernel.

\begin{table}
\centering
\caption{\small Table shows the rate of being identified as active by different algorithms in 50 repetitions of simulations with binary design matrix.}\label{tab:appendix sim 2 binary}
% \usepackage{multirow}
\begin{tabular}{|c|c|ccccccc|}
\hline
\multirow{2}{*}{Prior} & \multirow{2}{*}{Algorithm} & \multicolumn{7}{c|}{Proportion of Discovery}                             \\ \cline{3-9} 
                       &                            & Inactive Features         & $x_1$   & $x_2$   & $x_3$   & $x_4$  & $x_5$ & $x_6$   \\ \hline
                       & SSVGP                      & \multicolumn{1}{c|}{0.23} & 1  & 1  & 0.98    & 0.82    & 0.52 & 0.38 \\ \hline
\multirow{2}{*}{Exp(2)} & Algorithm 2 with $\alpha_{80}$ & \multicolumn{1}{c|}{0.20} & 1  & 1  & 0.98    & 0.80 & 0.34 & 0.22 \\ \cline{2-9} 
                       & Algorithm 2 with $\alpha_{90}$ & \multicolumn{1}{c|}{0.05} & 1  & 0.98  & 0.82 & 0.36 & 0.08 & 0.06 \\ \hline
\end{tabular}
\end{table}

\subsubsection{Simulation 3}
\medskip
The design matrix $\mathrm{X}$ of this example is generated by drawing a $500 \times 54$ matrix (i.e., $n=500,K=54$) of standard normal variables, and the mean of the dependent variables
is determined by
\begin{equation}
    E(y_i|\mathbf{x}_i)= \mu_i = \sin{x_{i, 1}} + \frac{3}{2} \cos{x_{i, 2}} + 2 \sin{x_{i, 3}}+\frac{5}{2} \cos{x_{i, 4}},\quad i=1,\ldots,500,
\end{equation}
where $\mathbf{x}_i$ is the feature vector of the $i_{th}$ observation, and $x_{i, k}$ represents the $k_{th}$ element in $\mathbf{x}_{i}$. Then the dependent variable $y_i$ is sampled from $\mathcal{N}(\mu_i, 1)$. In this simulation, we have 4 active features and 50 absolutely inactive features. 

Again, the simulation is repeated 50 times. Table \ref{tab:appendix sim 3} shows the proportion of discovery of the active features $x_1,\ldots,x_4$, and the inactive features from SSVGP and Algorithm 2 with Exp(2) prior for inverse length-scales. From the results, our proposed method can achieve even better performance than SSVGP by selecting an optimal threshold. 
\begin{table}
\centering
\caption{\small Table shows the rate of being identified as active by different algorithms in 50 repetitions of simulations.}\label{tab:appendix sim 3}
% \usepackage{multirow}
% \usepackage{multirow}
\begin{tabular}{|c|c|ccccc|}
\hline
\multirow{2}{*}{Prior} & \multirow{2}{*}{Algorithm} & \multicolumn{5}{c|}{Proportion of Discovery}               \\ \cline{3-7} 
                       &                            & Inactive Features         & $x_1$   & $x_2$   & $x_3$   & $x_4$   \\ \hline
                       & SSVGP                      & \multicolumn{1}{c|}{0.09} & 1  & 0.98 & 1  & 0.98 \\ \hline
\multirow{2}{*}{Exp(2)} & Algorithm 2 with $\alpha_{80}$ & \multicolumn{1}{c|}{0.16} & 1  & 1    & 1  & 1    \\ \cline{2-7} 
                       & Algorithm 2 with $\alpha_{90}$ & \multicolumn{1}{c|}{0.04} & 1  & 1    & 1  & 1    \\ \hline
\end{tabular}
\end{table}

\subsection{The Distribution of the Number of EEG Activated Locations}\label{sec: dist of activation}
\begin{figure*}
        \centering
        \begin{subfigure}[b]{0.48\textwidth}
            \centering
            \includegraphics[width=\textwidth]{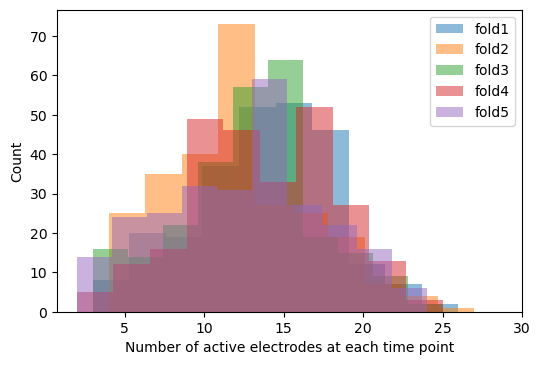}
            \caption[]%
            {{\small LGP.1 with $\alpha_{50}$}}    
            \label{fig: random_active_1}
        \end{subfigure}
        \hfill
        \begin{subfigure}[b]{0.48\textwidth}  
            \centering 
            \includegraphics[width=\textwidth]{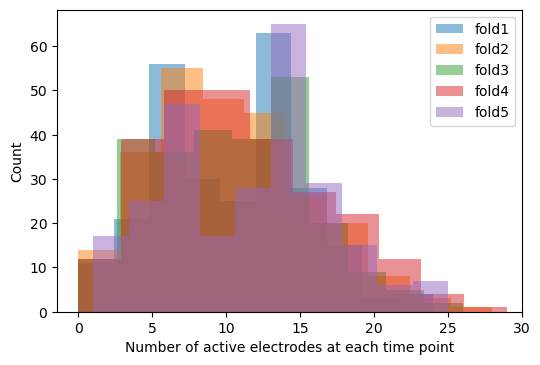}
            \caption[]%
            {{\small LGP.2 with $\alpha_{50}$}}    
            \label{fig: pca_active_1}
        \end{subfigure}
        
        \begin{subfigure}[b]{0.48\textwidth}
            \centering
            \includegraphics[width=\textwidth]{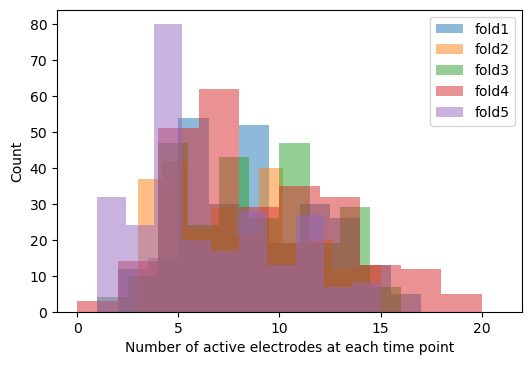}
            \caption[]%
            {{\small LGP.1 with $\alpha_{55}$}}    
            \label{fig: random_active}
        \end{subfigure}
        \hfill
        \begin{subfigure}[b]{0.48\textwidth}  
            \centering 
            \includegraphics[width=\textwidth]{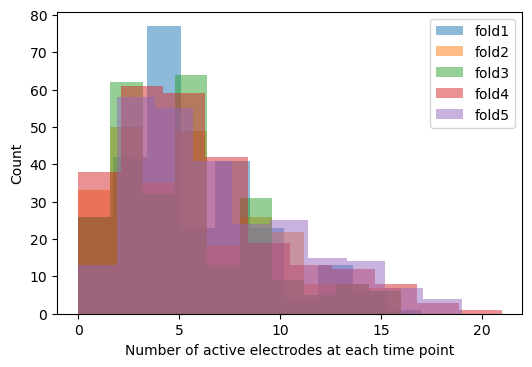}
            \caption[]%
            {{\small LGP.2 with $\alpha_{55}$}}    
            \label{fig: pca_active}
        \end{subfigure}
        
        \begin{subfigure}[b]{0.48\textwidth}
            \centering
            \includegraphics[width=\textwidth]{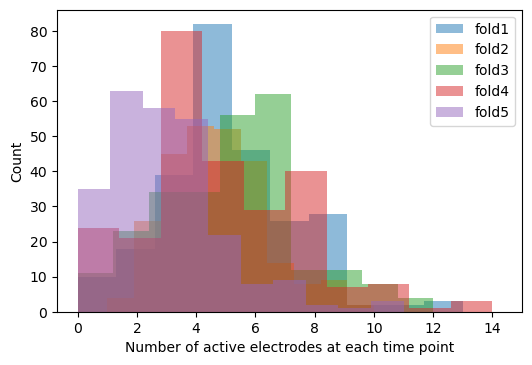}
            \caption[]%
            {{\small LGP.1 with $\alpha_{60}$}}    
            \label{fig: random_active_60}
        \end{subfigure}
        \hfill
        \begin{subfigure}[b]{0.48\textwidth}  
            \centering 
            \includegraphics[width=\textwidth]{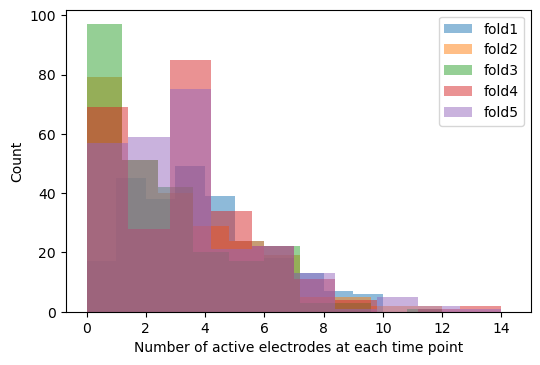}
            \caption[]%
            {{\small LGP.2 with $\alpha_{60}$}}    
            \label{fig: pca_active_60}
        \end{subfigure}

        \caption{\small The distribution of the number of EEG activated locations at each time point identified by (a) LGP.1 with $\alpha_{50}$, (b) LGP.2 with $\alpha_{50}$, (c) LGP.1 with $\alpha_{55}$, (d) LGP.2 with $\alpha_{55}$, (e) LGP.1 with $\alpha_{60}$, and (f) LGP.2 with $\alpha_{60}$ in a 5-fold cross-validation.}
        \label{fig: active}
    \end{figure*}

\medskip
Figure \ref{fig: active} shows the histograms of the number of activated locations at each time point identified by our proposed algorithms with different threshold in a 5-fold cross-validation. As the cut-off increases, the distributions move toward left. The Algorithm 2 tends to provide more sparse models. When $\alpha_{60}$ is used, it tends to over-sparsify the models to have only a few active features, especially for LGP.2 method.

\subsection{The EEG Activated Locations with $\alpha_{50}$}\label{sec: eeg_50}
\begin{figure*}
        \centering
        \begin{subfigure}[b]{0.8\textwidth}
            \centering
            \includegraphics[width=\textwidth]{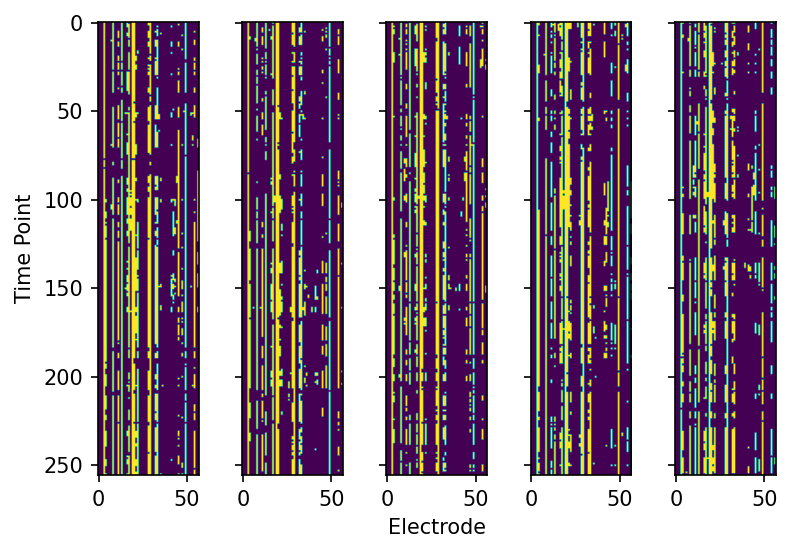}
            \caption[]%
            {{\small LGP.1 with $\alpha_{50}$}}    
            \label{fig: eeg_random_50}
        \end{subfigure}
        \vfill
        \begin{subfigure}[b]{0.8\textwidth}  
            \centering 
            \includegraphics[width=\textwidth]{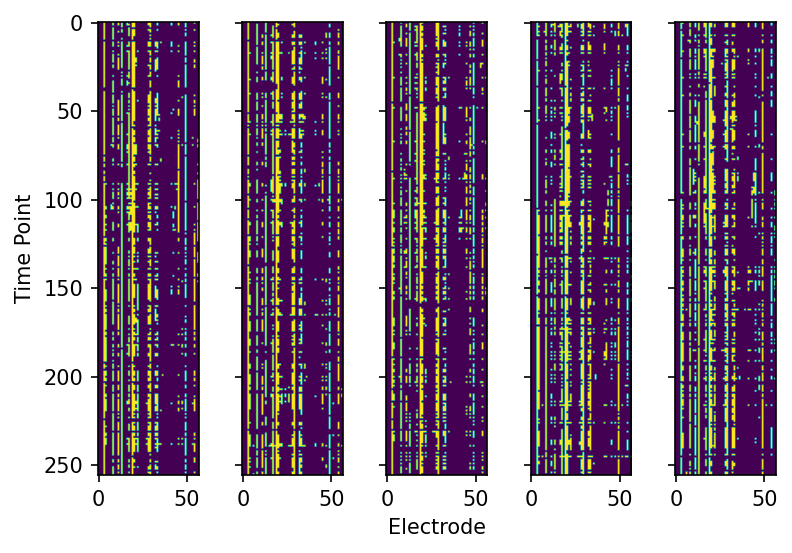}
            \caption[]%
            {{\small LGP.2 with $\alpha_{50}$}}    
            \label{fig: eeg_pca_50}
        \end{subfigure}
        \hfill

        \caption{\small EEG activated locations obtained from (a) LGP.1 with $\alpha_{50}$ and (b) LGP.2 with $\alpha_{50}$ in a 5-fold cross-validation. The black area indicates that the electrode at corresponding time point is not considered active by the algorithm. The plot from left to right in a sub-figure is obtained from cross-validation fold 1 to 5 respectively.}
        \label{fig: eeg_50}
    \end{figure*}

\medskip
Figure \ref{fig: eeg_50} shows EEG activated locations across time that obtained from LGP methods with $\alpha_{50}$ in a 5-fold cross-validation. The black area indicates that the location of electrode at corresponding time point is not considered active by the algorithm. The plot from left to right in a sub-figure is obtained from cross-validation fold 1 to 5 respectively.

\bibliographystyle{chicago}
\bibliography{ref}